# BRINGING PARTICLE PHYSICS INTO CLASSROOMS

K. CECIRE[1], I. MELO[2], B. TOMÁŠIK[3]

[1]*University of Notre Dame,* [2]*University of Žilina,* [3]*Matej Bel University*

E-mail: melo@fyzika.uniza.sk

*on behalf of the International Particle Physics Outreach Group*

Exciting scientific results such as the discovery of the Higgs boson offer a great opportunity to engage young people in particle physics. International Particle Physics Masterclasses highlight how high school students across the world can be exposed to real data from CERN's LHC accelerator in a stimulating and productive atmosphere in just a single day.

## INTRODUCTION

The term "masterclass" is familiar to millions worldwide; students often take part in masterclasses in the arts, whether they be music, visual arts, dance, or some other form. In these masterclasses, students learn about their artistic medium and improve their technique by intensive work under an expert "master." The greatest value is in the interaction between the master and the students where they learn much more than just improving the performance or project at hand.

International Masterclasses in particle physics [1,2] do much the same thing as masterclasses in the arts, but the medium and the master are different. The canvas for students in International Masterclasses is a set of event displays showing authentic data from actual particle physics experiments. To analyze these events, students interact with particle physicists, the masters. In the same way as in the arts, the students not only learn about the underlying physics but also about how to understand the behavior of the experimental instruments and how to get the most out of them. Since 2011, the instruments have become the four main detectors—ALICE, ATLAS, CMS, and LHCb—at CERN's Large Hadron Collider (LHC), and the masters have been physicists working on these experiments.

From their beginning as a local activity in the United Kingdom in the late 1990s to International Masterclasses today, masterclasses have evolved and grown. We will trace this growth and examine the progress of International Masterclasses worldwide and in Slovakia.

## HOW MASTERCLASSES WORK

An International Masterclass in particle physics is typically a one-day event at an institution such as a university or laboratory. Students will, in many cases, prepare beforehand in their schools with their physics teachers. This is done in the United States, for example, and U.S. masterclass leaders have observed that preparation helps students get the most out of the masterclass day.

Either with or without preparation, the masterclass day begins in the morning with an ice-breaker such as observation of a cloud chamber or e/m apparatus. Then students are treated to a short lesson on the Standard Model and experimental particle physics and a tour of an experimental physics facility. (It does not need to be a particle physics facility; what is really needed is an interesting experiment with an enthusiastic explainer to show the students around.)

Lunchtime can be part of the program. In many masterclasses, students eat with physicists and chat informally about everything from supersymmetry to favorite music.

Students get a second lesson on how to analyze the data. When protons collide in the LHC, the energy of the collisions can create relatively massive, unstable particles which decay promptly. The daughter particles from these decays are detected in the LHC detectors and can be used to understand the parent particles from which they came. Students visually study these decay events one-by-one using computer-rendered visualizations called "event displays". An example of such an event display is in Figure 1 below. After a presentation detailing how to access the event display, interpret what they see, and compile results, students have a good idea of the task ahead. Students begin to analyze data. As we expect, they run into problems, get confused, and seek help from the physicists. They really understand their task and, most importantly, how to see data as an "apprentice" physicist.

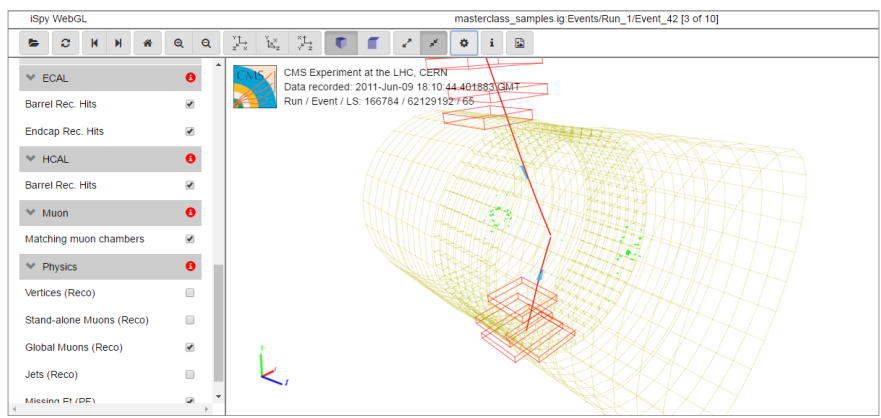

Figure 1. CMS event display as students might see it in a masterclass. This event shows a muon-antimuon pair.

Students work in pairs (two students at one computer) to analyze events to help and challenge each other. They rely on two-fold coincidence of opinions to have confidence in their analysis. For example, they may be looking for dilepton events which may be evidence of decays of Z bosons. Such events are most often electron-positron or muon-antimuon pairs. One student may see two electron-like tracks and conclude the parent particle may have been a Z. The student's partner may point out that both "electrons" appear to have positive charge, meaning, based on charge conservation, they could not as a pair have come from the decay of a neutral Z. Thus, the partners can check on each other and, when they cannot reconcile their opinions, seek assistance.

As the students work, physicists and Ph.D. students circulate and help students as needed. These are the interactions that resolve more difficult problems and lead to the deepest level of understanding. As in Figure 2, experts help students as they work by explaining their experience with data analysis, explaining behaviors of particles or of the detector, and posing challenging questions. Sometimes, they will lay out alternatives for the students, explain the physics behind those choices, and then leave to students to think it through.

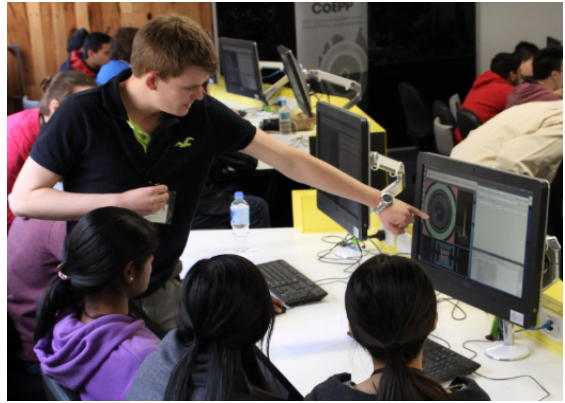

Figure 2. A physicist helps students in an ATLAS masterclass in Australia.

As the students make progress in their data analysis, they find that actual particle physics events do not behave in a "textbook" fashion. In learning from each other and from interaction with physicists, students find out that nature does not provide answers "from the back of the book," and that rational judgment is important trying to distinguish between signal and background or in matching real events to what is expected. They may even see two physicists disagree over the interpretation of a difficult event. Working out these difficulties and resolving a best approach to data analysis are part of how physicists write the analysis code to study millions of events and draw conclusions.

**HISTORY**

Particle physics masterclasses started quietly, with a conversation in the U.K. Institute of Physics in 1996 between Ken Long of Imperial College London and Roger Barlow of Manchester University [3]. They were looking for a new, more effective form of particle physics outreach and came up with an idea. Masterclasses were first done in the U.K. using real data from LEP, the Large Electron-Positron collider at CERN, which was then operating in the 27-kilometer-circumference tunnel which now houses the LHC. Student reaction was positive and others began to pick it up and develop the idea.

By the World Year of Physics, 2005, the LEP masterclasses were well established using data from the OPAL and DELPHI detectors. Erik Johansson of Stockholm University, who had contributed the DELPHI measurement, and Michael Kobel of Technical University Dresden took masterclasses to the next level by making them the cornerstone activity of the European Particle Physics Outreach Group (EPPOG, now IPPOG, the International Particle Physics Outreach Group) [4,5]. The new International Masterclasses featured institutions at multiple locations each doing the same analysis with high school students and then meeting in a CERN-moderated videoconference to discuss results [2]. There were 72 masterclasses that year in 18 countries for some 3,000 students [6]. It was a great beginning.

In 2006, a group of students did a masterclass at Brookhaven National Laboratory and joined the CERN videoconference the next day, beginning the U.S. participation and a partnership with QuarkNet. The internationalization of International Masterclasses did not stop there. Over time, International Masterclasses have spread to the Middle East, Asia, the Americas, Oceania, and Africa. Figure 3 shows how truly "international" the International Masterclasses have become.

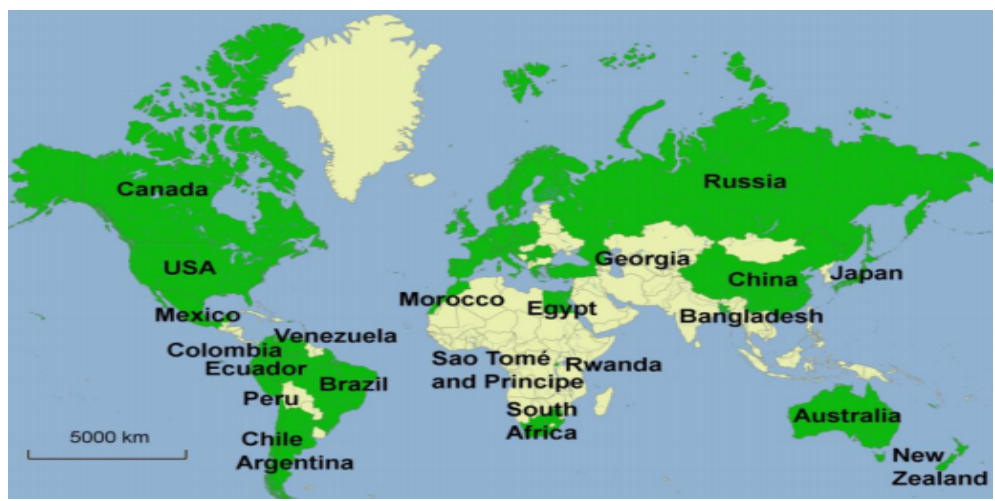

Figure 3. Participation by country (green colour) in International Masterclasses in 2017. European participation includes Austria, Belgium, Croatia, Cyprus, Czech Republic, Denmark, Finland, France, Germany, Greece, Hungary, Ireland, Italy, Montenegro, Netherlands, Norway, Poland, Portugal, Romania, Russia, Serbia, Slovakia, Slovenia, Spain, Sweden, Switzerland, Turkey, United Kingom. Not shown explicitly but participating is also Israel.

In 2011, International Masterclasses switched from LEP to all-LHC measurements. In 2013, Higgs candidate real events were added to the ATLAS and CMS measurements and by 2014 all four of the main LHC experiments were represented. By 2016, there were 213 masterclasses that year in 75 countries for some 13,000 students, with further growth expected with the compilation of statistics for 2017 [6].

**MASTERCLASSES IN SLOVAKIA**

Slovakia has been part of the event since 2005 when masterclasses turned international for the first time. Currently we organize masterclasses in universities in seven out of eight country's regions. It is interesting to note that only two of these universities offer undergraduate/graduate programs in particle physics and just four have particle physicists as staff members. The Masterclass team travels to the remaining universities to join forces with local physicists from other fields to assure that high school students can always listen and talk to experts in particle physics. Even though outside of their research scope, the management of these universities finds the event worth supporting – the high profile of International Masterclasses builds up their positive image and brings large numbers of prospective students into their auditoria (Figure 4) and computer rooms.

In total, close to 400 students attend International Masterclasses annually. In addition, one or two regional masterclasses are organized outside the international framework. Here the particle physicists travel to a high school far from any of the seven universities which host the international event, to give a chance to students who could not participate otherwise. The regional masterclass includes lectures and measurements but not the video-conference.

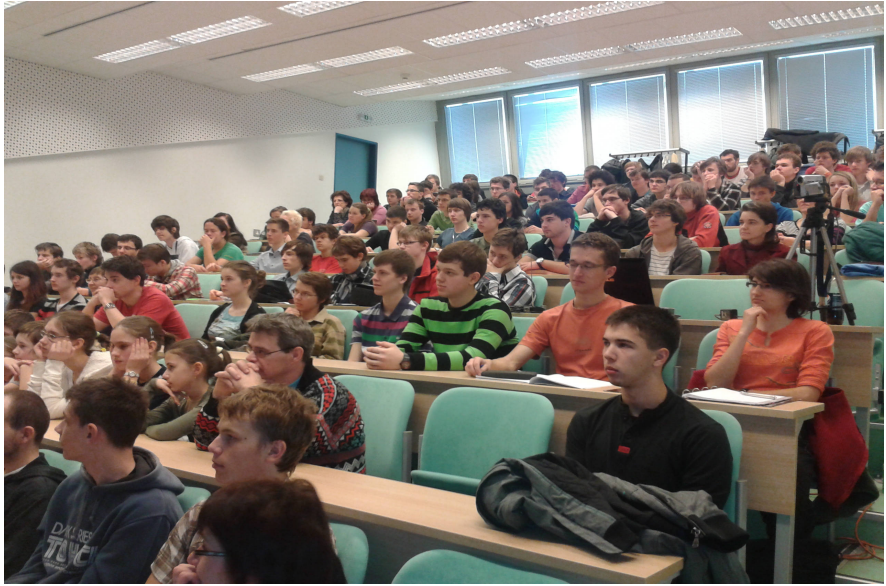

Figure 4. Masterclass at University of Žilina

Slovakia made important contributions to the progress of the International Masterclasses. For several years the video-conferences at the end of the day were technically supported by members of the Slovak-Caltech team based in Košice and CERN. Physicists from Košice and Prešov were among the first to develop astrophysical (2007) and cosmic ray (2011) variations of the traditional LEP/LHC Masterclass measurements. In 2007 the global evaluation of International Masterclasses (1482 questionnaires) was done at University of Žilina. In some years we motivated our students to write essays about their masterclass experiences and one of them, written by Košice Masterclass participant Slávka Marcinová, was published by [CERN Courier](#) in 2010 [7]. Starting in 2010 we organize annual Cascade projects competition, a follow-up to masterclasses for students who would like to know more about particle physics [8].

In 2013-2015 we conducted a survey based on 590 questionnaires (roughly 200 each year) filled out by Slovak masterclass participants at the end of the day. The average age of the participant was 16.8 years. Female students represented 39% of the total. As our survey shows (Table 1), 72% of participants are interested to pursue a career in natural sciences (including Informatics) or engineering programs. There is also a significant fraction (10%) interested in Medicine and 18% in social sciences and humanities (listed as Other in Table 1). We feel that it is of paramount importance that these future opinion makers in their field of interest have a better appreciation and understanding of the role of science in modern society from our program.

Table 1. Answers to the question "What would you like to study?"

| Physics | Math | Informatics | other Natural Sciences | Engineering | Medicine | Other |
|---|---|---|---|---|---|---|
| 25% | 12% | 17% | 9% | 9% | 10% | 18% |

Curiously, 28% of participants at the end of the day also answered that they were influenced by masterclasses in their choice of future studies. In Table 2 we show preferences the students indicated for a follow-up event which might enhance their knowledge of particle physics. A direct contact with an expert in the field and practical exercises are among the favourites.

Table 2. Answers to the question "What kind of follow-up event on this topic would you prefer?"

| Lecture by an expert | Film | Practical exercise/ Measurement | International competition - festival | International video-conference with experts |
|---|---|---|---|---|
| 38% | 27% | 34% | 19% | 20% |

In Table 3 we show preferences students had for individual parts of the programme. The highest positive rating was given to Lectures and Measurements (LHC data evaluation). On the other side, 63 out of 590 pointed to the Video-conference as the part which needs further attention from the organizers.

Table 3. Numbers of students who liked (second row) and did not like (third row) individual parts of the Masterclass programme. Multiple answers were allowed.

|  | Lectures | Visit to local laboratories | Practical exercise/measurement | Video-conference | Assistants | Venue | Lunch |
|---|---|---|---|---|---|---|---|
| Liked | 435 | 101 | 419 | 270 | 210 | 202 | 349 |
| Did not like | 36 | 16 | 24 | 63 | 10 | 25 | 37 |

In the overall evaluation of the International Masterclasses, 52% of the Slovak students rated the programme with the best mark "1", 31% with "2", 9% with "3", 2% with "4" and 5% chose the worst mark "5".

**MASTERCLASSES AS PART OF CURRICULA**

Over the years particle physics masterclasses took different forms addressing not just high-school students but various audiences including high-school teachers, the general public and students in developing countries. Other physics fields, in particular nuclear physics, astroparticle physics and cosmic rays physics are also using the masterclasses as a model for informal physics education. All these masterclass varieties are discussed in more detail in [2].

In the latest development, masterclasses are making their way into formal physics education as well. In Germany, in the state with the largest population (North Rhine Westfalia), particle physics became part of the mandatory curriculum and state-wide problems for the baccalaureat exam will be ready soon. Netzwerk Teilchenwelt is doing in-service teachers training [9] with new school material in 4 volumes and masterclass data analysis [10].

Australia aims to introduce the IPPOG Masterclasses programme as part of their formal science education in high-schools in New South Wales. The NSW Dept. of Education is beginning a trial of the online resource in 2017.

Masterclasses also offer an opportunity to develop a university level introductory particle physics course for engineering students. Such a course (optional in a curriculum) is offered to $1^{st}$ year students of the Master degree programme at the Faculty of Electrical Engineering at University of Žilina, Slovakia. Lectures (2 hours/week) include introduction to the Standard model of particle physics, accelerators and detectors while tutorials (2 hours/week) rely on CMS and ATLAS masterclass measurements with LHC data. The course was inspired by the high demand for skilled engineers at the largest particle physics laboratories and the natural interest some Engineering students have in modern physics.

**CONCLUSIONS**

International Particle Physics Masterclasses serve as an example that world-wide collaboration is essential not only for modern research itself but also for bringing its results into classrooms in a meaningful and inspiring way for high-school students. The programme continues to foster scientific culture within society in many countries while it is slowly aiming to find its place also in the formal science education system.